\documentclass[onecolumn,preprintnumbers,amsmath,amssymb]
{revtex4-2}
\usepackage{graphicx}
\usepackage[normalem]{ulem}
\usepackage[svgnames]{xcolor}
\usepackage{bm}
\usepackage{multirow}\usepackage{float}
\usepackage{titlesec}
\usepackage{xcolor}
\usepackage{amssymb}
\usepackage{amsmath}
\usepackage{float}
\begin{document}
\title{Thermal conductivity of Barium Bismuthate at low temperatures}

\author{A. \surname{Henriques}$^1$}

\author{D. M. N. \surname{Oliveira}$^1$}

\author{M. \surname{Baksi}$^2$}

\author{M. \surname{Naveed}$^1$}

\author{W. H. \surname{Brito}$^3$}

\author{J. \surname{Larrea-Jimenez}$^1$}

\author{D. \surname{Kumah}$^2$}

\author{S. \surname{Wirth}$^4$}

\author{V. \surname{Martelli}$^1$}
\email{valentina.martelli@usp.br} 

\affiliation{$^1$Laboratory for Quantum Matter under Extreme Conditions,  Institute of Physics, University of São Paulo, 05508-090, São Paulo, Brazil\\\\
$^2$Department of Physics, 169 Partners III 851, North Carolina State University,  USA\\\\
$^3$Department of Physics, Federal University of Minas Gerais, C. P. 702, Belo Horizonte, MG,  Brazil\\\\
$^4$Max-Planck Institute for Chemical Physics of Solids, Nöthnitzer Straße 40, Dresden, Germany
}

\date{\today} 
\begin{abstract}
The perovskite BaBiO$_3$ crystallizes in a cubic structure and undergoes structural transitions toward lower symmetry phases upon cooling. 
The two low-temperature monoclinic phases are insulating, and the origin of this unexpected non-metallic character has been under debate. Both monoclinic phases exhibit tilting and breathing distortions, which are connected with the insulating nature of this compound and may have important effects on phononic heat conductivity.

Here, we report the first thermal conductivity measurement, $\kappa(T)$, in pristine polycrystalline BaBiO$_3$ from 1.5 K to 310 K. At low and intermediate temperatures, we observe features reminiscent of a glass-like behavior, whereas at high-temperatures we find a downturn - typical of a crystalline solid. We compare our findings with available data of other recently investigated perovskite oxides displaying similar temperature dependence.

\end{abstract}
\maketitle

Barium bismuthate BaBiO$_{3}$ (BBO) has attracted considerable attention due to its intriguing phase diagram and electronic properties \cite{Sleight2015, Plumb2016}. BBO exhibits an insulating ground state, whereas a metallic character is expected in a simple ionic picture considering a semi-filled Bi-6s shell \cite{Bouwmeester2021}. Its cubic phase becomes superconducting upon hole-doping when either the Ba-site is replaced by Pb \cite{Sleight1975, Mattheiss1988} or the Bi-site is substituted by K \cite{Bazhirov2013}, while the mechanism at the base of the superconducting state is still an open question. More recently, a theoretical prediction of a Topological Insulating state (TI) in electron-doped BBO  re-ignited the interest in this compound \cite{Yan2013, Malyi2020}. Over the last two decades, the feasibility of such a TI phase has been a matter of intense debate: theoretical calculations \cite{Malyi2020, Zhang2017} indicated that the massive amount of electrons required to elevate its Fermi energy by $\sim$ 2 eV in the cubic phase leads to structural instabilities, preventing the achievement of the predicted TI phase. Reduction of the sample thickness was then used as a tuning parameter to suppress tilting and breathing distortions \cite{Kim2015, Inumaru2008, Zapf2019}, in the attempt to stabilize the cubic phase. Nevertheless, no metallicity was observed, leaving the persistent insulating state in BBO an open mystery.

Two possible mechanisms are currently proposed to be at the origin of the insulating state of BBO: an electrical charge disproportionation (CD) of Bi$^{+3}$ and  Bi$^{+5}$ ions and a bond disproportionation associated with the hybridization of Bi-6s and O-2p orbitals with a unique bismuth valence \cite{Bouwmeester2021}. Both come along with breathing and tilting distortions of the Oxygen-octahedra (Figure \ref{fig:1}(a)), causing a periodic distortion of the lattice. Recently, the gap formation was interpreted within a fractional CD, where the fluctuations are driven by the dynamical breathing modes of the octahedra \cite{Sarkar2021}. Recent DFT+U+V calculations highlighted the importance of considering both on-site and intersite electronic interactions for the accurate description of the breathing distortions \cite{Jang2023}, suggesting a complex interplay between the electronic and lattice degrees of freedom.

Bulk BBO crystallizes in a cubic structure above 800 K and undergoes several structural transitions toward lower symmetry as a function of temperature. The crystalline BBO presents a rhombohedral structure (space group R$\bar{3}$) in the $400$ K $<T<800$ K range and a monoclinic phase for $140$ K $<T<400$ K  dubbed monoclinic II (space group I2/m)   \cite{Kennedy2006}. It finally displays a transition at $T<140$ K  into a monoclinic I-phase (P2$_1$/n group) \cite{Kennedy2006}. 

In this work, we present and discuss the temperature-dependent thermal conductivity $\kappa(T)$ of polycrystalline BBO from 310~K down to 1.5~K, a range covering the monoclinic I to II structural transitions. Our results bring new experimental insight into the debate about the persistent insulating state and lattice distortions in BBO.

\section{METHODS}

A two-step solid-state reaction was employed to synthesize crystals of BaBiO$_3$: a 2:1 molar ratio of powders of BaCO$_3$ and Bi$_2$O$_3$ was mixed and ground, and the mixture was annealed in an alumina crucible at 800 $^{\circ}$C for 20 hours in flowing oxygen. The resulting BBO powder was re-ground for the recrystallization step as reported by Balandeh et al., who obtained single crystals of a few millimeters \cite{Balandeh2017}. 
The apparent density of our crystals estimated from the weight and volume is $\rho = (7.8\pm0.2)$ $\text{g/cm}^3 $, consistent with the reported values \cite{Balandeh2017, Degani1990}.

XRD data of powder BBO were acquired with a Rigaku Ultima III Diffractometer under $K_\alpha$-Cu wavelength and angular step of 0.02$^{\circ}$ in the interval $15^{\circ} < 2\theta < 108^{\circ}$. Back-scattering Laue diffraction was employed to verify the crystallographic alignment. AC resistivity measurements using the four-probe method were carried out down to 200 K; electrical contacts were prepared using 50 $\mu$m-thick copper wires and silver paste. A standard one-heater-two-thermometers method was employed for measuring thermal conductivity, $\kappa(T)$, in the 1.5 - 310 K temperature range in a home-built setup inside a customized probe fitting a low-temperature stage. Thermal conductivity is calculated as $\kappa (T) = P/(g\Delta T)$ where $P$ is the power that establishes a temperature gradient $\Delta T$ along the sample and  $g = A/\ell$ is the specimen geometrical factor (cross-section area over the distance between the thermometers).

\section{RESULTS AND DISCUSSION}

\begin{figure}[ht]
\includegraphics[width=17cm]{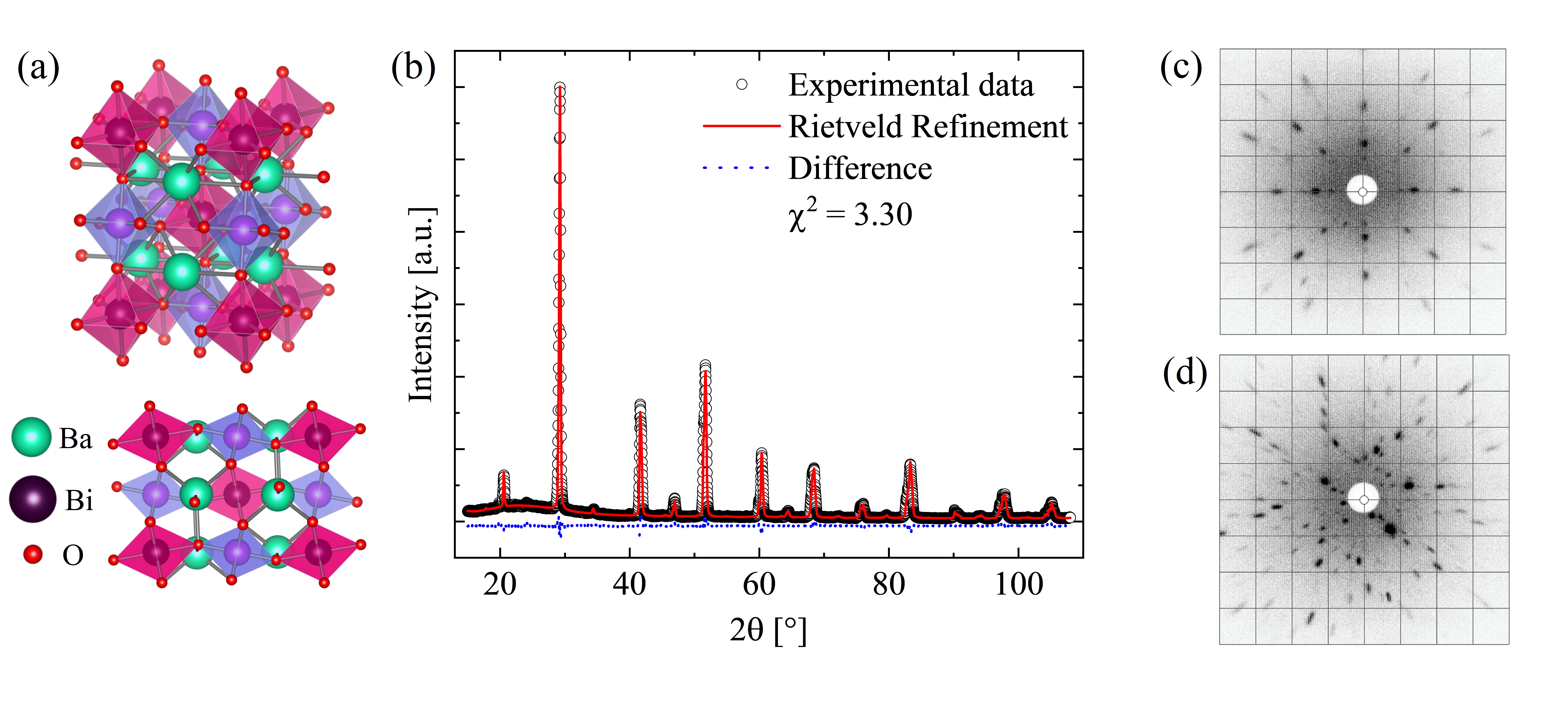} 
\caption{(a) 3D Structural model of the monoclinic-II phase of BaBiO$_3$ together with a 2D projection depicting the breathing and tilting distortions at different BiO$_6$ sites (b) XRD data and Rietveld refinement. (c) X-ray back-scattering Laue image of the grown crystal parallel to the [001] direction at a given spot of the specimen's surface. (d) Same as in (c) but taken at a different spot of the specimen's surface.}
\label{fig:1}
\end{figure}

XRD patterns were analyzed by using the GSAS-II software \cite{GSASII} as shown in Figure \ref{fig:1}(b). Inside the experimental resolution, Rietveld refinement shows one single phase with space group I2/m for the monoclinic II phase. The lattice parameters were found to be $a=6.1864$ \AA, $b = 6.1403$ \AA,  $c=8.6736$ \AA and $\beta = 90.06^{\circ}$, presenting a good agreement when compared with previously reported values (see Table \ref{tab1}). By transverse Laue x-ray scans, we found a spatial dependence on the diffraction spots formation - see Figure  \ref{fig:1}(c) and (d). At some positions, well-oriented grains were found but their non-homogeneity across the sample highlights the presence of coarse-grained crystallites, as the observed spots are not uniformly arranged in a concentric ring but concentrated into separate regions. 

Based on the (x-ray) illuminated area, the crystallite sizes were estimated to be in the order of $\sim 0.75$ mm$^2$. In this work, as the transport properties are investigated in samples with a length of the order of 5 mm, the specimens are polycrystals with domain sizes of the order of the crystallites. 

\begin{figure}[ht]
\includegraphics[width=9cm]{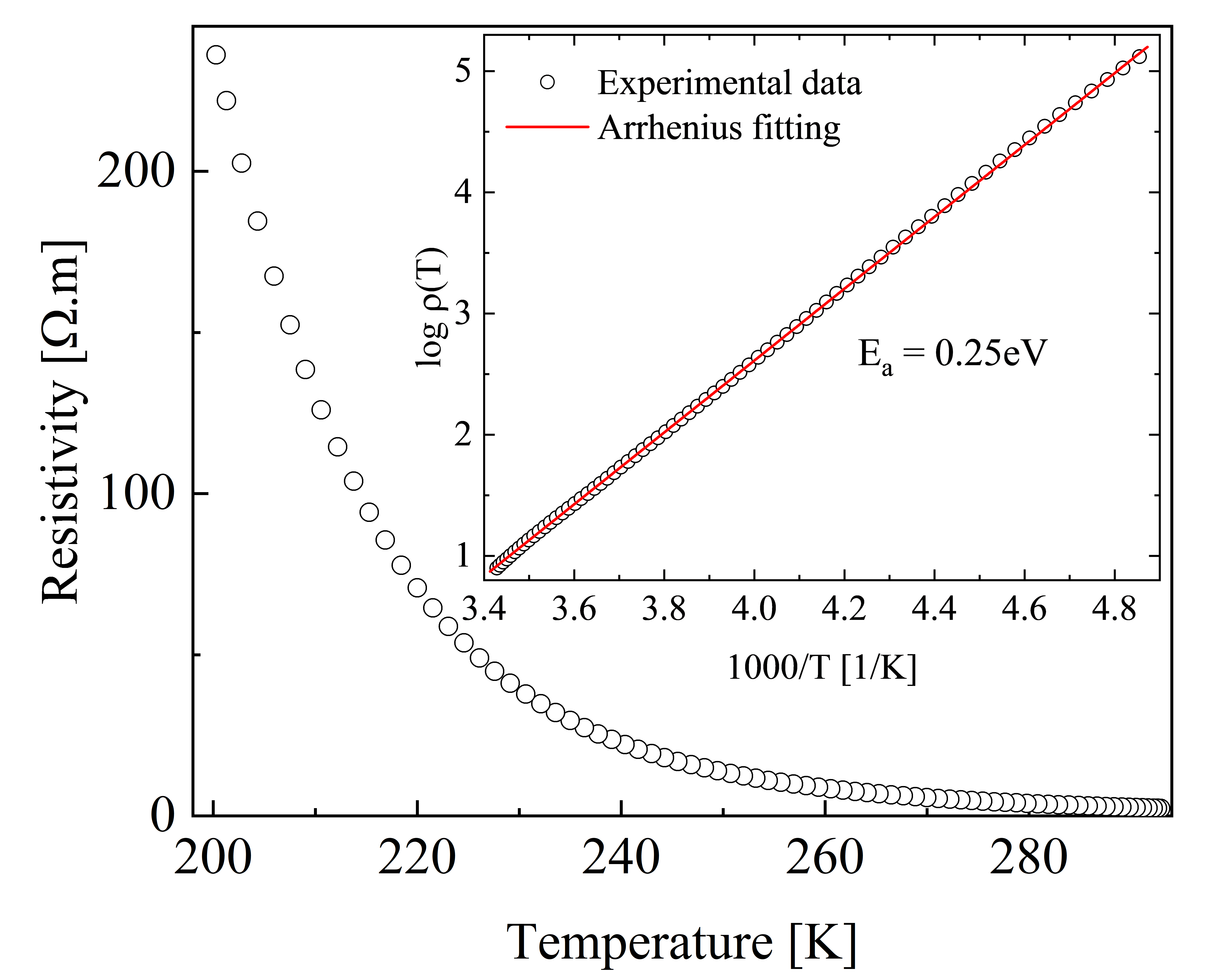} 
\caption{Electrical resistivity of BaBiO$_3$. The inset shows the activation energy $E_a = 0.25$ eV as determined by fitting the Arrhenius equation of the log-linearized data.}
\label{fig:resitivity}
\end{figure}

Analysis of electrical resistivity using the Arrhenius equation  $\rho(T) \sim \exp(E_a/k_B T) $ yields a good description of the semiconducting temperature dependence of resistivity, as shown in Figure \ref{fig:resitivity}. The activation energy of $E_a = 0.25$ eV agrees with previous literature studies \cite{Kim1995, Sleight2015, Kumar2016}. Overall, the investigated structural and electric transport properties point to samples with comparable quality to those reported in the literature.

\begin{table}[t]
\centering
\caption{Lattice parameters of the monoclinic II phase of BaBiO$_3$ at ambient temperature obtained from Rietveld Refinement and comparison to literature.}
\label{tab1}
\begin{ruledtabular}
\begin{tabular}{@{}llllll@{}}
\textbf{Reference} & \textbf{$a$} (\AA) & \textbf{$b$} (\AA) & \textbf{$c$} (\AA) & \textbf{$\beta$ ($^{\circ}$)} & Phase \\ \hline

Kennedy (2006) \cite{Kennedy2006} & 6.18505 & 6.13219 & 8.6585 & 90.229 & I2/m \\
Chaillout (1985) \cite{Chaillout1985} & 6.1721 & 6.1301 & 8.6617 & 90.048 & I2/m \\
Zhou (2004) \cite{Zhou2004} & 6.19125 & 6.15264 & 8.69177 & 90.1115 & I2/m \\
Yamaguchi (2005) \cite{Yamaguchi2005} & 6.188 & 6.141 & 8.675 & 90.16 & I2/m \\
Pei (1989) \cite{Pei1989} & 6.1863(1) & 6.1406(1) & 8.6723(1) & 90.164(2) & I2/m \\
Foyevtsov (2019) \cite{Foyevtsov2019} & 6.1903(27) & 6.1471(27) & 8.6819(33) & 90.0764(47) & I2/m \\

This work & 6.1864 & 6.1403 & 8.6736 & 90.06 & I2/m \\

\end{tabular}
\end{ruledtabular}
\end{table}

\begin{figure}[ht]
\includegraphics[width=17cm]{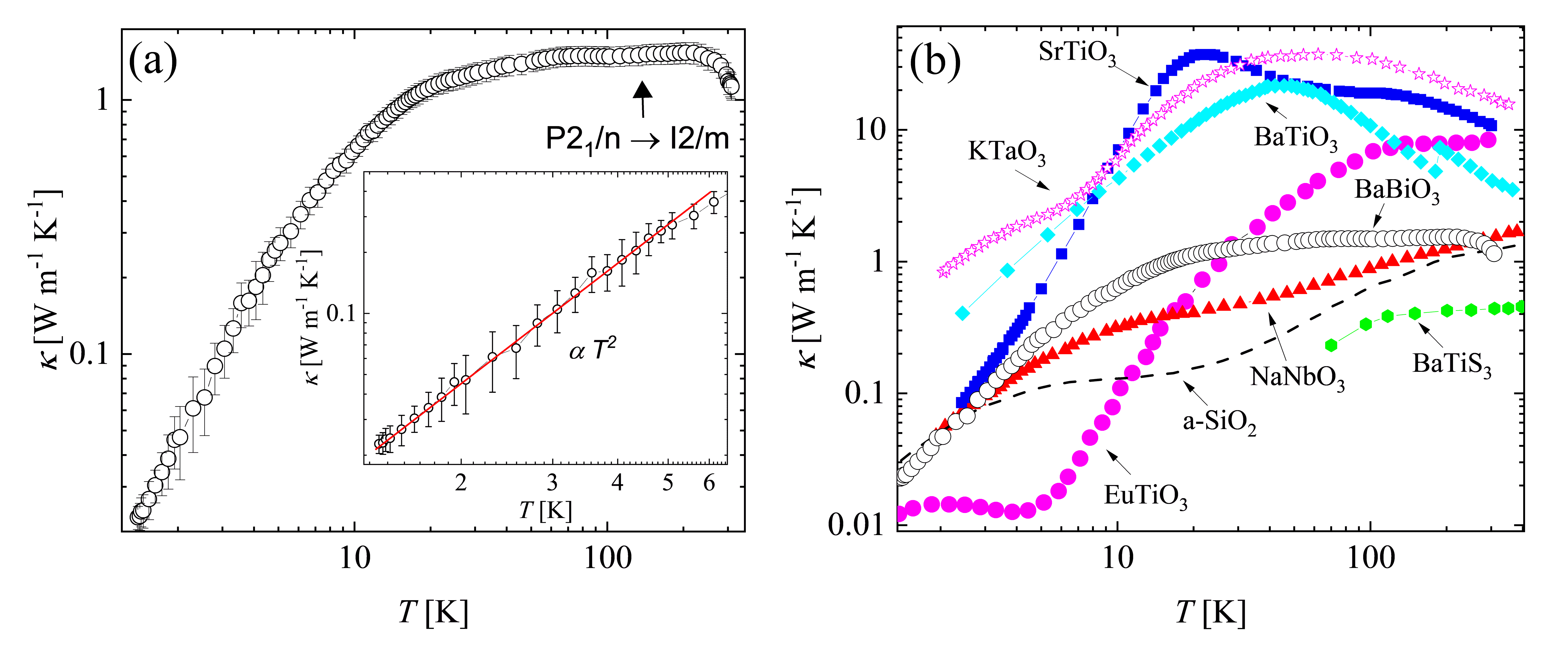} 
\caption{Thermal conductivity of BaBiO$_3$ and its comparison to various complex oxides and amorphous silica. (a) Thermal conductivity of BaBiO$_3$ in the 1.5 K to 310 K range. The inset shows the $T^2$ behavior at low temperatures. (b) $\kappa(T)$ of some related (mostly) perovskite systems. Data from SrTiO$_3$ \cite{Martelli2018}, a-SiO$_2$ \cite{Cahill1992}, BaTiS$_3$ \cite{Sun2020}, EuTiO$_3$ \cite{Jaoui2023} and others \cite{Tachibana2008}.}
\label{fig:kappaBBO}
\end{figure}

Figure \ref{fig:kappaBBO}(a) shows $\kappa(T)$ between 1.5 K and 310 K, which, to the best of our knowledge, has not been reported before. For completeness, we note that a single room-temperature value in the order of 1 W m$^{-1}$ K$^{-1}$ was mentioned with no further details in an early work \cite{Ponnambalam1998}.  The uncertainties on $\kappa(T)$ were calculated to be about 9\%, with the main source of uncertainty being the determination of the geometrical factor. We now proceed to the discussion of $\kappa(T)$ in the three typical temperature ranges.

\emph{At low temperatures} ($T<6$ K), $\kappa(T)$ shows a $T^\delta$ power-law with $\delta \sim 2.0$, as reported in the inset of Figure \ref{fig:kappaBBO}(a). At this temperature range, called as ballistic regime, a typical $T^3$ power law is usually observed in single crystals and in polycrystals, with a mean-free path that in the former case can approach the sample size and in the latter is limited by the crystallite size \cite{Vandersande1986}. Extended or point defects do not usually account for the observed $~T^2$ behavior due to the large phonons wavelength. Scattering by free electrons can also be discarded because of the semiconducting character of BaBiO$_3$. What then leads to a $\kappa(T)\sim T^2$ at low temperatures? One possible trivial explanation might reside in the presence of a large number of dislocations, which can lead to a similar power-law as shown in a study of LiF \cite{Sproull1959}. Another possibility might come from having a glass-like thermal conductivity, i.e., a behavior that mimics a two-level system that was called into question to explain the $T^2$ behavior in glasses and amorphous systems \cite{Anderson1971, Phillips1972, Jackle1972}. 

\emph{At intermediate temperatures} (at a fraction of the Debye temperature $\sim 0.1\Theta_D\sim 25~$ K \cite{Kuentzler1991}), the lack of a prominent peak in thermal conductivity is evident. Instead, an extended temperature range of approximately constant thermal conductivity  spans from $\sim 63$ K to $235$ K. Noticeably, a consistent correlation between the onset of the plateau in thermal conductivity of glass-like systems and their Boson peak temperature $T_{\text{peak}}$ in specific heat was clearly pointed out \cite{Hu2022}. For BBO, we have estimated $T_{\text{peak}} \sim 15$ K \cite{Kuentzler1991}, but the plateau shows up only above $\sim 60$ K. We finally observe that  the monoclinic-I to monoclinic-II transition (see arrow in Fig. \ref{fig:kappaBBO}(a)) happens within the plateau region, but well above the expected peak temperature ($\sim 0.1\Theta_D$); as in this work we are studying a polycrystal, we cannot draw an unambiguous conclusion \textit{i)} on the relation between the structural transition and the plateau extended in a wide range and \textit{ii)} whether the structural transition is in this case detectable by thermal conductivity measurements - as, e.g., observed in the sister-compound BaTiO$_3$ (see Figure \ref{fig:kappaBBO}(b)). It is important to remark that, although dislocations might bring a $T^2$ power law at low temperatures, they do not necessarily suppress the peak in $\kappa(T)$ \cite{Sproull1959}.

\emph{At high temperatures} ($T>200~$ K), $\kappa(T)$ shows a downturn up to the highest investigated temperature ($\sim$ 310 K). It is worth commenting that this feature is clearly displayed even before applying the usual radiative correction.  Despite $\kappa(T)$ in BBO displaying glass-like features at lower temperatures, in this region it surprisingly decreases as a Debye crystal but faster than $T^{-1}$. That downturn often does not appear for other compounds with glass-like thermal conductivity, Figure \ref{fig:kappaBBO}(b). We argue that this might be related to the high-temperature monoclinic phase, while the features observed at lower temperatures are linked to the monoclinic-I structural phase. 

The behavior of $\kappa(T)$ in BaBiO$_3$ at low and intermediate temperatures might suggest a glass-like thermal conductivity, which has been reported in other crystalline compounds \cite{Cohn1999, Tachibana2015, Salamatov2017}. Figure \ref{fig:kappaBBO}(b) shows our data and the thermal conductivity of other selected perovskites and amorphous silica. NaNbO$_3$ and BaTiS$_3$ were reported to display lattice glass-like thermal conductivity \cite{Tachibana2008, Sun2020}. In those systems, ferro/antiferroelectric phase coexistence and sub-THz frequency atomic tunnelling states that mimic a two-level system are suggested to be the mechanisms that drive the glass-like thermal conductivity, respectively. Recently, EuTiO$_3$ has been shown to exhibit glass-like thermal conductivity driven by superexchange and valence fluctuations \cite{Jaoui2023}. In the case of BBO, the origin of the glass-like behavior might be related to the dynamics of octahedra distortions, remaining a subject for future investigation at this stage. 

We end our discussion by comparing - see Table \ref{tab2} - a few relevant values related to the thermal properties of BBO and another complex oxide, SrTiO$_3$.  At room-temperature, the magnitude of thermal conductivity is ten times less in BaBiO$_3$  while $v_L$ (longitudinal sound velocity), $\Theta_D$, and $T_M$ (melting temperature) scale approximately as a factor 2. Further investigation on the thermal diffusivity in BBO and its consistency with theories and experimental evidence on its lower bound \cite{Martelli2018,  Behnia2019, Mousatov2020, Martelli2021} may be an insightful future step to quantitatively advance this discussion. 

\begin{table}[ht]
\centering
\caption{Thermal conductivity at room temperature $\kappa_{300K}$, longitudinal sound velocity $v_L$, Debye temperature $\Theta_D$ and melting temperature $T_M$ of BaBiO$_3$ calculated using data available in \cite{Gao2020, Klinkova1999, Kuentzler1991} and of SrTiO$_3$ \cite{Martelli2018, Poirier1989}.}
\label{tab2}
\begin{ruledtabular}
\begin{tabular}{@{}lllll@{}}
\textbf{Compound} & $\kappa_{300K}$ [W m$^{-1}$ K$^{-1}$] & $v_L$ [km/s] & $\Theta_D$ [K] & $T_M$ [K]\\ \hline

SrTiO$_3$ & 10.8 & 7.87 & 397 & 2213 \\
BaBiO$_3$ & 1.20 & 4.40 & 229 & 1313 \\

\end{tabular}
\end{ruledtabular}
\end{table}

\section{CONCLUSIONS}

BaBiO$_3$ polycrystals were obtained in a similar quality with respect to what was reported in the literature. The order of magnitude of barium bismuthate thermal conductivity was confirmed to be about 1 W m$^{-1}$ K$^{-1}$. Temperature-dependent measurements show features in $\kappa(T)$ that deviate from a typical insulating crystal, particularly the absence of thermal conductivity peak at $\sim 0.1\Theta_D$ and the $T^2$ dependence at low temperatures. Further investigations in BaBiO$_3$ single crystals are necessary to shed light on the origin of the observed $\kappa(T)$.

\begin{acknowledgements}
VM, AH, DMNO, MN acknowledge the support of  São Paulo Research Foundation (FAPESP) grant number 2018/19420-3, 2022/01742-0, 2021/00625-7, 2022/03262-5 respectively. VM and DK acknowledge the University Global Partnership Network (UGPN) Research Collaboration Fund (RCF). JLJ acknowledges support of FAPESP (2018/08845-3) and CNPq-PQ2 (310065/2021-6). WHB acknowledges CNPq (402919/2021-1). We thank A. C. Franco of the Multi-user Laboratory of Crystallography at IF-USP for support during the XRD measurements. 
\end{acknowledgements}

\bibliography{references}

\end{document}